\documentclass{jpsj3}
\usepackage{bm}
\title{Theory of the $\beta$-type Organic Superconductivity under Uniaxial Compression}
\author{Takeo Suzuki, Seiichiro Onari, Hiroshi Ito, and Yukio Tanaka}
\inst{\address{Department of Applied Physics, Nagoya University, Chikusa, Nagoya 464-8603, Japan}}

\abst{
We study theoretically the shift of the superconducting transition
 temperature ($T_c$) under uniaxial compression in 
$\beta$-type organic superconductors, $\beta$-(BEDT-TTF)$_{2}$I$_{3}$ and
$\beta$-(BDA-TTP)$_{2}$X[X=SbF$_{6}$,AsF$_6$], in order to clarify the electron 
correlation, the spin frustration, and the effect of dimerization. The transfer integrals 
are calculated by the extended H\"{u}ckel method assuming the uniaxial strain, and the 
superconducting state mediated by the spin fluctuation is solved using Eliashberg's equation 
with the fluctuation-exchange approximation. The calculation is carried out on both 
the dimerized (one-band) and nondimerized (two-band) Hubbard models. 
We have found that (i) the behavior of $T_c$ in $\beta$-(BEDT-TTF)$_{2}$I$_{3}$ with a
stronger dimerization is well reproduced by the dimer model, while that
in weakly dimerized 
$\beta$-BDA-TTP salts is rather well reproduced by the two-band model, and (ii) the competition 
between the spin frustration and the effect induced by the fluctuation is important in 
these materials, which causes the nonmonotonic shift of $T_c$ against uniaxial compression.
}

\kword{superconductivity, organic superconductor, uniaxial compression, spin frustration, 
spin fluctuation}
\begin{document}
\maketitle

%\preprint{APS/123-QED}

%It is always \today, today,
%but any date may be explicitly specified

%Use showkeys class option if keyword
%display desired
\section{Introduction}
\begin{figure}[hbtp]
 \begin{center}
 \includegraphics[width=7.5cm]{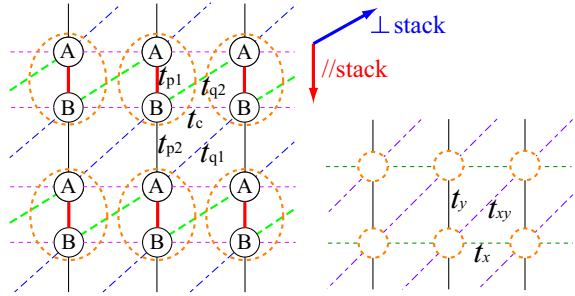}
 \caption{(Color online) Lattice structure on a two-band model (left) and the dimer model
  (right) for $\beta$-type salts. Donor molecules locate at A and B sites, and form dimers, as
  depicted by the dotted ellipsoids.}
\label{lattice}
 \end{center}
\end{figure}

The superconductivity of the $\kappa$-type BEDT-TTF salts has fascinated
experimental and theoretical researchers with a high transition temperature
$T_c$ over $10$ K.\cite{Urayama} In the $\kappa$-type BEDT-TTF salts, donors 
are considered to form dimers.\cite{Kino-Fukuyama} Then, the split antibonding 
HOMO band is half-filled with the effective on-site Coulomb interaction $(U)$ 
close to the bandwidth. It is considered to be a strongly correlated electron system, where
the superconductivity is mediated by the antiferromagnetic spin fluctuation.
\cite{Kino,Kondo,Powell,Kuroki,Watanabe}  The spin fluctuation is suppressed by the geometrical 
spin frustration since the donors locate on an anisotropic triangular
lattice. The effect becomes strongest on a regular triangular
lattice, which is considered to be realized in
$\kappa$-(BEDT-TTF)$_2$Cu$_2$(CN)$_3$.\cite{Komatsu,Shimizu} 
The organic superconductors with $\beta$-type donor arrangement also
exhibit a   
relatively high $T_c$ reaching 8 K.  As shown in Fig. \ref{lattice}, donors stack 
unidirectionally and dimerize on a triangular lattice. However, the dimerization of 
the molecules is weaker than that of the $\kappa$-type salts. Then, as an effective 
model for the $\beta$-type salt, not only the dimer model but also the nondimerized 
original two-band model should be taken into account. 

Uniaxial compression is a powerful technique for studying the strongly
correlated electron systems since transfer integrals can be controlled
selectively.\cite{Maesato1} The spin frustration and the electron correlation are 
controlled by uniaxial compression. In fact, the shift of $T_c$ under the uniaxial
compression in $\kappa$-(BEDT-TTF)$_2$Cu(NCS)$_2$ can be explained by
the term of the 
spin frustration on an anisotropic triangular lattice composed of dimerized 
molecules.\cite{Maesato2}  Our group reported the shift of $T_c$ in $\beta$-type salts, 
$\beta$-(BDA-TTP)$_2$X [X=SbF$_6$, AsF$_6$], under uniaxial compression parallel 
and perpendicular to the molecular stack direction.\cite{Ito1}  Under the compression 
perpendicular to the stack, $T_c$ shows a nonmonotonic behavior, once increases and 
reaches a maximum and decreases subsequently, as a function of the applied piston pressure. 
On the basis of the dimer model of the anisotropic triangular lattice, this behavior has been 
attributed to the result of the competition of the enhancements of the spin fluctuation and 
the spin frustration effect. However, the data for 
$\beta$-(BDA-TTP)$_2$AsF$_6$ was not well reproduced. Furthermore, for 
$\beta$-(BEDT-TTF)$_2$I$_3$, $T_c$ exhibits nonmonotonic behaviors for both 
compression directions, in strong contrast to that of the BDA-TTP salts.\cite{Ito2}

In order to clarify the behavior of $T_c$ under uniaxial compression and obtain insight into 
the superconductivity of $\beta$-type salts, we study theoretically the superconductivity 
mediated by the spin fluctuation on both the original two-band Hubbard model and the 
dimerized Hubbard model using the fluctuation exchange (FLEX) 
approximation \cite{Bickers1,Bickers2} for $\beta$-(BEDT-TTF)$_{2}$I$_{3}$ and
$\beta$-(BDA-TTP)$_{2}$X[X=SbF$_{6}$,AsF$_6$].  
Measurements of STM and specific heat 
are consistent with the $d$-wave symmetry of BDA-TTP salts 
\cite{Nomura,Shimojo}. 
It is noted that the zero bias conductance peak 
due to the Andreev bound state \cite{Nomura1,Kashiwaya} 
has been reported very recently \cite{Nomura2}. 
Thus, it is natural to  employ the spin fluctuation theory, which can realize a $d$-wave gap function beyond the BCS theory. 

Although some of the results have been 
reported in our previous papers \cite{Ito1,Suzuki,Ito2}, in this paper, we
note the difference in the strength of the dimerization among
the three materials, which is evaluated by the ratio of the transfer
integrals along the stacking direction, namely, $t_{p1}$ and $t_{p2}$,
as shown in
Fig. \ref{lattice}. We have found that the dimer model is suitable to the strongly
dimerized $\beta$-(BEDT-TTF)$_2$I$_3$ among the three materials, $t_{p1}/t_{p2}=2.99$, while
the original two-band model is suitable for the weakly dimerized $\beta$-
(BDA-TTP)$_2$AsF$_6$ among the three materials, $t_{p1}/t_{p2}=2.02$. 
We have also found that the nonmonotonic behavior of $T_c$ is due to the
competition between the spin frustration and the effect of the self-energy
induced by the spin fluctuation. 

%%%%%%%%%%
%%%%%%%%%%

\section{Formulation}

In the following, we introduce the models that we employ in this study. Putting a 
donor molecule into a site on a lattice, the crystal structure of $\beta$-type salts 
is modeled by a triangular lattice with A and B sublattices, as shown in Fig. \ref{lattice}.
Here, we distort the lattice in order to make the reciprocal lattice
rectangular. 
Then, the direction perpendicular to the donor stack direction is distorted, as illustrated 
in Fig. \ref{lattice}. 
However, this distortion makes no changes in physical quantities, such as $T_c$, 
because the topology of the lattice is invariant.
According to the dimer model, the sublattices are dimerized, 
and the original two-band model is reduced to the one-band model (dimer model), as shown in 
Fig. \ref{lattice}. In the dimer model, hopping integrals are given by $t_x=-t_{\rm
q2}/2+t_c$, $t_{xy}=-t_{\rm q1}/2$, and $t_y=-t_{\rm p2}/2$ using the hopping integrals 
in the original two-band model.\cite{Kontani2003} Hopping integrals under uniaxial 
compression are also estimated by the extended H\"{u}ckel method \cite{Mori1984} assuming 
uniform displacements of molecules from the structure at ambient
pressure.\cite{Ito1} 
The validity of the Huckel method as a first approximation is implied because it can 
predict the round Fermi surface, which is consistent with the
Shubnikov-de Haas (SdH) or de
Haas-van Alphen (dHvA) measurements.\cite{Kang,Beckmann}
The uniaxial strain under uniaxial compression is assumed to be 0.3\%/kbar according to 
the results of X-ray measurements.\cite{Tanino} We also assume that the encapsulation of 
the sample with epoxy, which is prepared for uniaxial compression, induces an isotropic 
pressure of 1kbar in the sample. 
We show the obtained transfer integrals under uniaxial compression in
$\beta$-type salts in Table \ref{hop-table}.

\begin{table*}[bt]
\caption{Transfer integrals under uniaxial compression in the
 $\beta$-type salts, where unit of transfer integrals is $10^{-2}$eV.}
\begin{center}
%\scalebox{0.65}[0.65]{
\begin{tabular}{|c|c|c|c|c|c|c|} \hline
&Pressure [kbar]& $t_c$ & $t_{p1}$ & $t_{p2}$ & $t_{q1}$ & $t_{q2}$\\
\hline
$\beta$-(BEDT-TTF)$_{2}$I$_{3}$ & 0 & -5.22& 24.5 & 8.23 & 12.6 & 6.91
 \\
\cline{2-7}
& //stack 5 & -4.75 & 26.8 & 9.16 & 13.0 & 7.07\\
\cline{2-7}
 & $\perp$stack 5 & -5.98 & 24.0 & 8.15 & 13.1 & 7.25\\
\hline
$\beta$-(BDA-TTP)$_{2}$SbF$_{6}$ & 0 & -0.45 & 15.0 & 6.43 & 8.34 &
 9.04\\
\cline{2-7}
& //stack 5 & -0.02 & 16.3 & 7.12 & 8.58 & 9.23\\
\cline{2-7}
 & $\perp$stack 5 & -0.85 & 14.8 & 6.38 & 8.68 & 9.60\\
\hline
$\beta$-(BDA-TTP)$_{2}$AsF$_{6}$ & 0 & -0.36 & 15.5 & 7.74 & 8.52 &
 9.79\\
\cline{2-7}
& //stack 5 & 0.07 & 16.8 & 8.45 & 8.72 & 9.99\\
\cline{2-7}
 & $\perp$stack 5 & -0.78 & 15.2 & 7.69 & 8.83 & 10.5\\
\hline
\end{tabular}
\end{center}

\label{hop-table}
\end{table*}

In this paper, we employ both the two-band and one-band Hubbard models with the repulsion 
$U$ on a triangular lattice,
\begin{eqnarray}
{\cal H}&=&\sum_{\langle i,j\rangle,\sigma}
\sum_{\alpha,\beta}^{\rm A,B}t_{ij}\left(c_{i\sigma}^{\alpha\dagger}c_{j\sigma}^{\beta}+{\rm H.c.}\right)\nonumber\\
& &+ U\sum_{i}\sum_{\alpha}^{\rm A,B}n_{i\uparrow}^{\alpha}n_{i\downarrow}^{\alpha},
\end{eqnarray}
in standard notations and calculate $T_c$ by solving Eliashberg's equation 
within the FLEX approximation.\cite{koikegami,kontani} In the two-band model, 
Green's function $G$, the self-energy $\Sigma$, the susceptibility $\chi$, and the gap function
$\phi$ are all $2 \times 2$ matrices, such as $G_{\alpha\beta}(k)$,
where $\alpha,\beta$=A, B and $k\equiv({\bm k},i\omega_n)$ with $\omega_n=(2n-1)\pi T$ 
being the Matsubara frequency for fermions.

The self-energy is given by
\begin{equation}
\Sigma_{\alpha\beta}(k)=\frac{T}{N}\sum_qG_{\alpha\beta}(k-q)V_{\alpha\beta}^{(1)}(q),\label{equbegin}
\end{equation}
where the FLEX effective interaction $V(q)$ for the self-energy is
\begin{eqnarray}
V_{\alpha\beta}(q)&=&\frac{3}{2}U^2\!\!\left[\frac{\chi^{{\rm
					      irr}}(q)}{1-U\chi^{{\rm
					      irr}}(q)}\right]_{\alpha\beta}\!\!\!+\frac{1}{2}U^2\!\!\left[\frac{\chi^{{\rm irr}}(q)}{1+U\chi^{{\rm irr}}(q)}\right]_{\alpha\beta}\\\nonumber
&&-U^2\chi^{{\rm irr}}_{\alpha\beta}(q)
\end{eqnarray}
with the irreducible susceptibility
\begin{equation}
\chi_{\alpha\beta}^{{\rm irr}}(q)=-\frac{T}{N}\sum_{k}G_{\alpha\beta}(k+q)G_{\beta\alpha}(k).
\end{equation}
Here, we denote $q\equiv({\bm q},i\epsilon_l)$  with $\epsilon_l=2\pi lT$ being the 
Matsubara frequency for bosons, and $N$ as the number of ${\bm k}$-points 
on a mesh.

With Dyson's equation,
\begin{equation}
\left[G(k)^{-1}\right]_{\alpha\beta}=\left[G^{0}(k)^{-1}\right]_{\alpha\beta}-\Sigma_{\alpha\beta}(k),\label{equend}
\end{equation}
where $G^{0}$ is the bare Green's function
$G_{\alpha\beta}^{0}(k) = 
[(i\omega_n+\mu-\varepsilon_{\bm k}^{0})^{-1}]_{\alpha\beta}$
with $\varepsilon_{\bm k}^{0}$ being the bare energy, 
we have solved eqs. (\ref{equbegin})-(\ref{equend}) 
self-consistently to obtain Green's functions under FLEX interactions.

Then, $T_c$ is obtained from Eliashberg's equation for the 
spin-singlet pairing,
\begin{eqnarray}
\lambda\phi_{\alpha\beta}(k)&=&-\frac{T}{N}\sum_{k'}\sum_{\alpha',\beta'}^{\rm A,B}V_{\alpha\beta}^{\rm sin}(k-k')\nonumber\\
&&\times G_{\alpha\alpha'}(k')G_{\beta\beta'}(-k')\phi_{\alpha'\beta'}(k')
\label{eliashberg},
\end{eqnarray}
where $\phi$ is the gap function, 
and the spin singlet pairing interaction $V^{\rm sin}(q)$ is given as
\begin{eqnarray}
V_{\alpha\beta}^{\rm sin}(q)&=&\frac{3}{2}U^2\!\!\left[\frac{\chi^{{\rm
						irr}}(q)}{1-U\chi^{{\rm
						irr}}(q)}\right]_{\alpha\beta}\!\!\!-\frac{1}{2}U^2\!\!\left[\frac{\chi^{{\rm
						irr}}(q)}{1+U\chi^{{\rm
						irr}}(q)}\right]_{\alpha\beta}\\\nonumber
&&+U\delta_{\alpha\beta}
.
\end{eqnarray}
From eq. (\ref{eliashberg}), $T_c$ is determined as the temperature at which 
the maximum eigenvalue $\lambda$ becomes unity. However, because the performance of the 
computer is not sufficient for deducing $T_c$ directly, we evaluate $\lambda$ at a constant 
temperature of $T/t_{q1}=0.01$ as a function of the piston pressure for the uniaxial 
compression. The larger value of $\lambda$ corresponds to the higher $T_c$.

At this temperature, the $\bm{k}$-dependence of the spin susceptibility $\chi^s$
is expressed by the diagonalized component
\begin{equation}
\chi_{\rm s}=\frac{\chi_{{\rm AA}}^{\rm s}+\chi_{\rm BB}^{\rm s}}{2}+\sqrt{\left[\frac{\chi_{\rm AA}^{\rm s}-\chi_{\rm BB}^{\rm s}}{2}\right]^{2}+\left|\chi_{\rm AB}^{\rm s}\right|^{2}},
\end{equation}
where the matrix of the spin susceptibility is 
\begin{equation}
\chi_{\alpha\beta}^{\rm s}({\bm k},0)=\left[\frac{\chi^{{\rm irr}}({\bm k},0)}{1-U\chi^{{\rm irr}}({\bm k},0)}\right]_{\alpha\beta}.
\end{equation}
Throughout this study, we consider $t_{\rm q1}$ under ambient pressure as a unit
of energy, $U=10$ on the two-band model and $U=2t_{\rm p1}$ on the dimer
model. The system with $U=10$ corresponds to the strongly correlated electron system. 
In the numerical calculation, we consider $N=128\times 128$ ${\bm k}$-point meshes and the
Matsubara frequencies from $\epsilon_n$ from $-(2N_c-1)\pi T$ to
$(2N_c-1)\pi T$ with $N_c=4096$.

\section{Result}

\subsection{$\beta$-(BEDT-TTF)$_{2}$I$_{3}$}

%%%%%%%%%%
\begin{figure}[btp]
 \begin{center}
 \includegraphics[width=5cm]{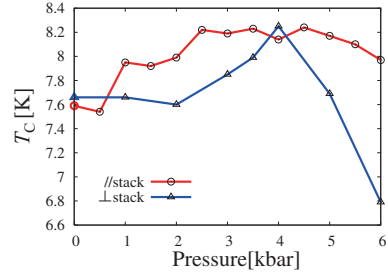}
 \caption{(Color online) Experimental data of $T_c$ under uniaxial compression in
$\beta$-(BEDT-TTF)$_{2}$I$_{3}$ defined by the midpoint of the transition.\cite{Ito2}}
\label{3}
 \end{center}
\end{figure}%
%%%%%%%%%%

We begin with $\beta$-(BEDT-TTF)$_{2}$I$_{3}$ in which the dimerization
is the strongest among 
the three salts. Although it is known that the $T_c$ of $\beta$-(BEDT-TTF)$_{2}$I$_{3}$ jumps 
from 1.5K at ambient pressure up to 8K at a pressure of 1kbar,\cite{Murata} only the 
8K superconductivity phase is observed under epoxy encapsulation because of the slight 
pressure induced by the epoxy.\cite{Ito2}  Figure \ref{3} shows that the
experimental 
$T_c$ increases with uniaxial compression along both the //stack and $\perp$stack 
directions up to $4$kbar, and decreases in further piston pressures.\cite{Ito2}
It is very interesting that $T_c$ shows a nonmonotonic behavior against the
uniaxial compression.

%%%%%%%%%%
\begin{figure}[hbtp]
 \begin{center}
 \includegraphics[width=7.5cm]{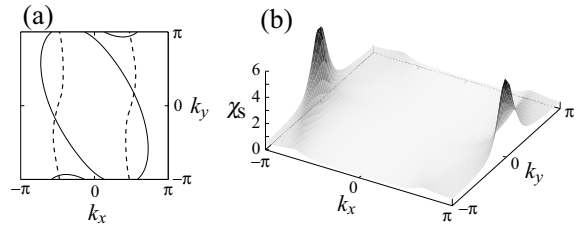}
 \caption{(a) Fermi surface (solid line) and nodes of gap function
  (dotted line) and (b) $\chi_s$ obtained using the two-band model for
  $\beta$-(BEDT-TTF)$_{2}$I$_{3}$ at ambient pressure.}
\label{2}
 \end{center}
\end{figure}%
%%%%%%%%%%

We first examine the gap function and the Fermi surface. The gap
function obtained using eq. 
(\ref{eliashberg}) is a $d$-wave that changes sign four times on the Fermi surface, as shown 
in Fig. \ref{2}. The Fermi surface obtained using the two-band model is closed owing to the 
periodicity of the Brillouin zone (BZ), which is consistent with the previous 
result\cite{Mori} and dHvA measurements\cite{Kang,Beckmann}.
The Fermi surface in the two-band model is almost the same as that in the dimer model.
The peak of $\chi_{\rm s}$ is observed at the wave number $(\pi,0)$. The gap function changes sign 
along the nesting vector $\bm{Q}=(\pi,0)$. We consider that $\beta$-(BEDT-TTF)$_{2}$I$_{3}$ 
is close to the SDW phase because the peak value of $\chi_{\rm s}$ is large ($\sim 6$).

%%%%%%%%%%
\begin{figure}[hbtp]
 \begin{center}
 \includegraphics[width=8.5cm]{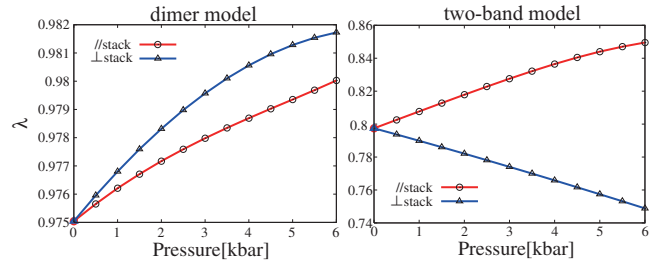}
 \caption{(Color online) $\lambda$ under uniaxial compression on the dimer model 
 (left panel) and two-band model (right panel) for $\beta$-(BEDT-TTF)$_{2}$I$_{3}$.}
\label{4}
 \end{center}
\end{figure}%
%%%%%%%%%%

Next, we see the shift of $\lambda$, {\it i.e.}, $T_c$ under uniaxial compression.
Assuming a linear response to the uniaxial strain, the experimental
results show an increase in $T_c$ under low piston pressures along both directions. 
Here, the absolute value of $\lambda$ is not important. Thus, we focus on the shift of $\lambda$ around
ambient pressure.
In the case of the dimer model, 
$\lambda$ increases with uniaxial compression along both the //stack and $\perp$stack 
directions at low pressures, as shown in Fig. \ref{4}. The trend of the increasing $\lambda$ along both compression directions agrees with the experimental 
result at low pressures, although the larger $\lambda$ under compression perpendicular to 
the stacks does not match with the order of the experimental $T_c$.

On the other hand, in the case of the two-band model, $\lambda$ increases with uniaxial 
compression along the //stack direction, while $\lambda$ decreases along
the $\perp$stack direction. 
This result does not match the experimental results. Thus, the dimer model is more 
suitable for $\beta$-(BEDT-TTF)$_{2}$I$_{3}$, which reproduces the shift of $T_c$ at  
low pressures, than the two-band model. However, the nonmonotonic behavior of $T_c$, 
{\it i.e.}, the subsequent decrease in $T_c$ at high pressures, can be reproduced by neither 
model.

\subsection{$\beta$-(BDA-TTP)$_{2}$SbF$_{6}$}

%%%%%%%%%%
\begin{figure}[btp]
 \begin{center}
 \includegraphics[width=5cm]{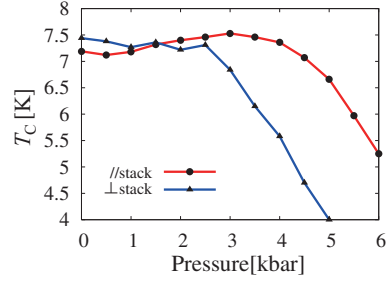}
 \caption{(Color online) Experimental data of $T_c$ under uniaxial compression in
$\beta$-(BDA-TTP)$_{2}$SbF$_{6}$ defined by the midpoint of the
  transition.\cite{Ito1}}
\label{fig5}
 \end{center}
\end{figure}%
%%%%%%%%%%

Next, we move on to $\beta$-(BDA-TTP)$_{2}$SbF$_{6}$. Under uniaxial compression along 
the //stack direction, experimental data of $T_c$ increases with
pressure, reaching a maximum at 
$3$kbar, and decreases in further pressures, as shown in
Fig. \ref{fig5}. On the other hand, along the $\perp$stack 
direction, $T_c$ decreases monotonically with pressure.\cite{Ito1}

%%%%%%%%%%
\begin{figure}[hbtp]
 \begin{center}
 \includegraphics[width=7.5cm]{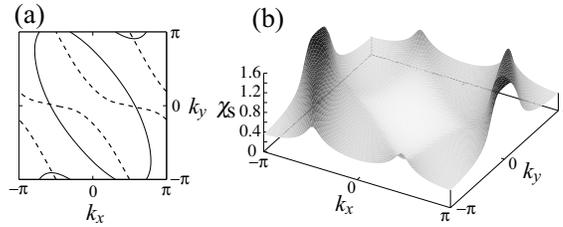}
 \caption{(a) Fermi surface (solid line) and nodes of gap function
  (dotted line) and (b) $\chi_s$ obtained using the two-band model for
$\beta$-(BDA-TTP)$_{2}$SbF$_{6}$ at ambient pressure.}
\label{5}
 \end{center}
\end{figure}%
%%%%%%%%%%

The gap function is also calculated to be a $d$-wave, as shown in Fig. \ref{5}(a).
The obtained Fermi surface and symmetry of the gap function are consistent with
previous calculations \cite{Yamada,Nonoyama} and STM measurements \cite{Nomura},
indicating the four-fold symmetry of the gap function. The $d$-wave is mediated by 
the spin fluctuation induced by the nesting vector. Thus, it is also
reasonable to apply the spin fluctuation theory to this system. As shown
in Fig. \ref{5}(b), the peak of $\chi_{\rm s}$ is found at the wave number ($\frac{9}{10}\pi$,$\frac{3}{4}\pi$); however, the peak value of 
$\chi_{\rm s}$ is small ($\sim 1.4$).

%%%%%%%%%%
\begin{figure}[hbtp]
 \begin{center}
 \includegraphics[width=8.5cm]{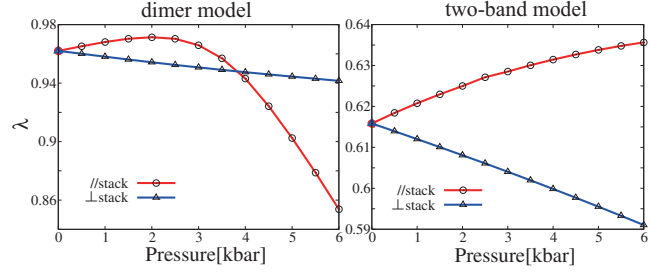}
 \caption{(Color online) $\lambda$ under uniaxial compression on the dimer model (left
  panel) and the two-band model (right panel) for $\beta$-(BDA-TTP)$_{2}$SbF$_{6}$.}\label{fig7}
 \end{center}
\end{figure}%
%%%%%%%%%%

Then, we observe the shift of $\lambda$ under uniaxial compression in Fig. \ref{fig7}. In the case of the 
dimer model, $\lambda$ increases with pressure reaching a maximum at $2$kbar, and decreases 
in further pressures under uniaxial compression along the //stack direction, while along the
$\perp$stack direction, $\lambda$ decreases monotonically with
increasing pressure. This behavior is 
consistent with the experimental results. On the other hand, in the case of 
the two-band model, $\lambda$ increases monotonically with pressure under uniaxial 
compression along the //stack direction, while $\lambda$ decreases monotonically under 
uniaxial compression along the $\perp$stack direction.

In $\beta$-(BDA-TTP)$_{2}$SbF$_{6}$, a maximum $\lambda$ value is obtained on the dimer 
model, which is not obtained in $\beta$-(BEDT-TTF)$_{2}$I$_{3}$. In the following, 
we discuss the origin of the nonmonotonic behavior of $\lambda$ under uniaxial 
compression calculated in the dimer model. 

First, we focus on $\langle U^2\chi_{\rm S}\rangle_{\rm B}$ that is the mean value of 
$U^2\chi_{\rm s}$ within the BZ, which is a significant term in Eliashberg's equation 
(\ref{eliashberg}). In the case of the dimer model, the on-site Coulomb interaction 
$U= 2t_{\rm p1}$ depends on the uniaxial compression. Thus, the value of $U^{2}$ 
in the effective interaction is also important when we discuss the
uniaxial compression 
dependence of $\lambda$.
%%%%%%%%%%
\begin{figure}[hbtp]
 \begin{center}
 \includegraphics[width=5cm]{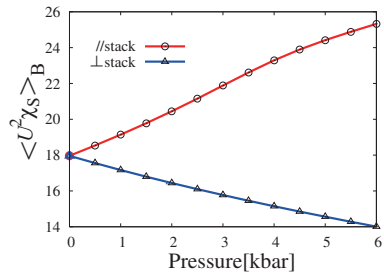}
 \caption{(Color online) $\langle U^{2}\chi_{\rm S}\rangle_{\rm B}$ under uniaxial 
  compression on the dimer model for $\beta$-(BDA-TTP)$_{2}$SbF$_{6}$.}
\label{8}
 \end{center}
\end{figure}%
%%%%%%%%%%
As shown in Fig. \ref{8}, under uniaxial compression along the //stack direction, 
$\langle U^{2}\chi_{\rm s}\rangle_{\rm B}$ increases monotonically with pressure, while 
$\langle U^{2}\chi_{\rm s}\rangle_{\rm B}$ decreases under uniaxial compression along 
the $\perp$stack direction. Here, we note the geometrical spin frustration induced by the
triangular lattice. We denote $t'(=t_y)$ as the hopping integral along the dimerized direction 
and $t(=(t_x+t_{xy})/2)$ as the mean value of other hopping integrals in the dimer model.
In this material, $t'/t<1$ is satisfied at ambient pressure, and then the geometrical spin
frustration becomes strong under uniaxial compression along the //stack direction,
while it becomes weak under uniaxial compression along the $\perp$stack direction, which is 
consistent with the results of $\langle U^2\chi_{\rm s}\rangle_{\rm B}$.

%%%%%%%%%%
\begin{figure}[hbtp]
 \begin{center}
 \includegraphics[width=5cm]{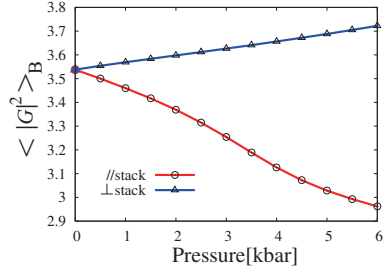}
 \caption{(Color online) $\langle |G|^{2}\rangle_{\rm B}$ under uniaxial compression 
 on the dimer model for $\beta$-(BDA-TTP)$_{2}$SbF$_{6}$.}
\label{9}
 \end{center}
\end{figure}%
%%%%%%%%%%

Next, we show in Fig. \ref{9} the mean value of $|G|^{2}$ in the BZ, which is another 
important term in eq. (\ref{eliashberg}). Under uniaxial compression
along the //stack 
direction, the value of $\langle |G|^{2}\rangle_{\rm B}$ decreases monotonically with increasing
pressure, while that increases monotonically along the $\perp$stack direction. Under 
uniaxial compression along the //stack direction, the increase in 
$\langle U^{2}\chi_{\rm s}\rangle_{\rm B}$ strongly competes with the decrease in 
$\langle |G|^{2}\rangle_{\rm B}$. Thus, the nonmonotonic 
behavior reaching a maximum at the value of $\lambda$ is considered to be caused by such 
competition.

%%%%%%%%%%
\begin{figure}[hbtp]
 \begin{center}
 \includegraphics[width=5cm]{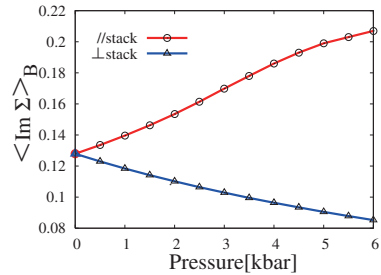}
 \caption{(Color online) $\langle |{\rm Im}\Sigma|\rangle_{\rm B}$ under uniaxial 
 compression on the dimer model for $\beta$-(BDA-TTP)$_{2}$SbF$_{6}$.
}
\label{10}
 \end{center}
\end{figure}%
%%%%%%%%%%

Now, we consider the origin of the uniaxial compression dependence in 
$\langle |G|^{2}\rangle_{\rm B}$. $G$ is suppressed by the imaginary part of self-energy 
Im$\Sigma$. In Fig. \ref{10}, we show the compression dependence of 
$\langle |{\rm Im}\Sigma|\rangle_{\rm B}$, which is the mean value of
$|{\rm Im}\Sigma|$ in the BZ.
Under uniaxial compression along the //stack direction, the value of 
$\langle|{\rm Im}\Sigma|\rangle_{\rm B}$ increases monotonically with pressure, while 
it decreases monotonically along the $\perp$stack direction. These results are consistent 
with the results of $\langle |G|^{2}\rangle_{\rm B}$. Thus, under uniaxial compression 
along the //stack direction, the increase in the value of 
$\langle|{\rm Im}\Sigma|\rangle_{\rm B}$ reduces the value of 
$\langle |G|^{2}\rangle_{\rm B}$, which competes with the increase in the value of 
$\langle U^{2}\chi_{\rm s}\rangle_{\rm B}$

Compared with the experimental results, the low-pressure behavior of $T_c$ can be 
reproduced by both the dimer and two-band models. We observe that the dimer model is 
more suitable than the two-band model, because the maximum of $T_c$ can be reproduced 
by the dimer model.

\subsection{$\beta$-(BDA-TTP)$_{2}$AsF$_{6}$}

%%%%%%%%%%
\begin{figure}[hbtp]
 \begin{center}
 \includegraphics[width=5cm]{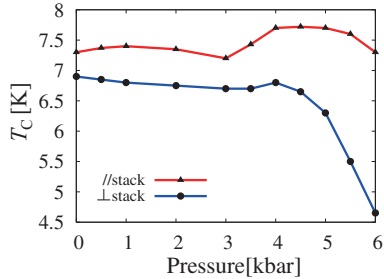}
 \caption{(Color online) Experimental data of $T_c$ under uniaxial compression in
 $\beta$-(BDA-TTP)$_{2}$AsF$_{6}$ defined by the midpoint of the transition.\cite{Ito1}}
\label{12}
 \end{center}
\end{figure}%
%%%%%%%%%%

Finally, we move on to $\beta$-(BDA-TTP)$_{2}$AsF$_{6}$. As shown in Fig. \ref{12}, under 
uniaxial compression along the //stack direction, the experimental data of $T_c$ is almost 
constant up to $3$kbar and increases with pressure, reaching a maximum at $4.5$kbar, and 
decreases in further pressures. On the other hand, along $\perp$stack direction, 
$T_c$ decreases monotonically with increasing pressure.\cite{Ito1}

%%%%%%%%%%
\begin{figure}[hbtp]
 \begin{center}
 \includegraphics[width=7.5cm]{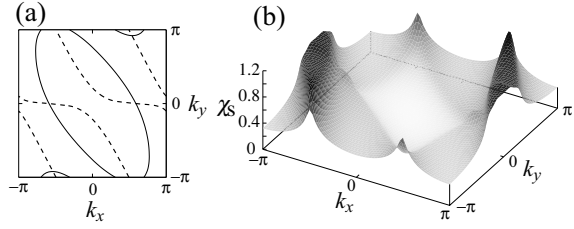}
 \caption{(a) Fermi surface (solid line) and nodes of gap function
  (dotted line) and (b) $\chi_s$ obtained using the two-band model in
  $\beta$-(BDA-TTP)$_{2}$AsF$_{6}$ at ambient pressure.}
\label{11}
 \end{center}
\end{figure}%
%%%%%%%%%%

As shown in Fig. \ref{11}, the calculated gap function is also a $d$-wave, and the shape of the 
Fermi surface is consistent with previous results \cite{Yamada} and STM measurements.
\cite{Nomura} The peak of $\chi_{\rm s}$ is found at the wave number ($\pi$,$\frac{2}{5}\pi$), 
but the peak value of $\chi_{\rm s}$ is small ($\sim 1.2$). We can assume that this material 
is far from the SDW phase owing to the small value of $\chi_{\rm s}$.

%%%%%%%%%%
\begin{figure}[hbtp]
 \begin{center}
 \includegraphics[width=8.5cm]{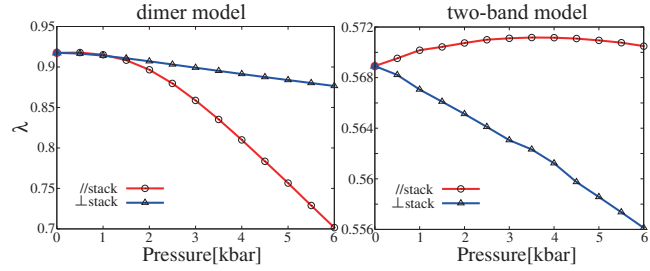}
 \caption{(Color online) $\lambda$ under uniaxial compression on the dimer model (left
  panel) and two-band model (right panel) for $\beta$-(BDA-TTP)$_{2}$AsF$_{6}$.}
\label{13}
 \end{center}
\end{figure}%
%%%%%%%%%%

Figure \ref{13} shows that in the dimer model, $\lambda$ increases only
very slightly up to 1 kbar and decreases monotonically with increasing
pressure under uniaxial compression along the //stack direction and $\lambda$ also 
decreases monotonically along $\perp$stack direction, which does not fit with the 
experimental results.\cite{Ito1} On the other hand, in the two-band model, $\lambda$ increases 
with pressure, reaching a maximum at $3.5$kbar, and decreases in further pressures under 
uniaxial compression along the //stack direction, while $\lambda$ decreases monotonically 
along the $\perp$stack direction. These behaviors of $T_c$ in the two-band model are 
consistent with those observed by experiment. Thus, the two-band model is more suitable for 
reproducing the behavior of $T_c$ in $\beta$-(BDA-TTP)$_{2}$AsF$_{6}$.

%%%%%%%%%%
\begin{figure}[hbtp]
 \begin{center}
 \includegraphics[width=5cm]{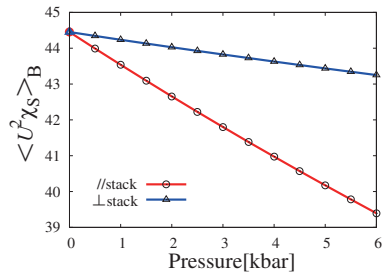}
 \caption{(Color online) $\langle U^{2}\chi_{\rm S}\rangle_{\rm B}$
  under uniaxial compression on the two-band model for $\beta$-(BDA-TTP)$_{2}$AsF$_{6}$.}
\label{14}
 \end{center}
\end{figure}%
%%%%%%%%%%

%%%%%%%%%%
\begin{figure}[hbtp]
 \begin{center}
 \includegraphics[width=5cm]{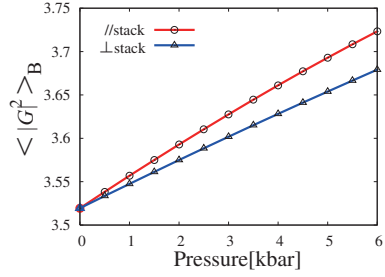}
 \caption{(Color online) $\langle |G|^{2}\rangle_{\rm B}$ under uniaxial compression 
 on the two-band model for $\beta$-(BDA-TTP)$_{2}$AsF$_{6}$.}
\label{15}
 \end{center}
\end{figure}%
%%%%%%%%%%

Hereafter, we discuss the origin of the nonmonotonic behavior of $\lambda$ in the 
two-band model. As shown in Fig. \ref{14}, in the two-band model, 
$\langle U^{2}\chi_{\rm S}\rangle_{\rm B}$ decreases with increasing pressure under uniaxial 
compression along both the //stack and $\perp$stack directions.
In Fig. \ref{15}, we show the result of $\langle|G|^{2}\rangle_{\rm B}$ that is 
the mean value of the largest eigenvalue of $|G|^2$ in the BZ.
Under uniaxial compression along both the //stack and $\perp$stack
directions, the value of 
$\langle |G|^{2}\rangle_{\rm B}$ increases monotonically with pressure. Thus, 
under uniaxial compression along the //stack direction, the decrease in 
$\langle U^{2}\chi_{\rm s}\rangle_{\rm B}$ and the increase in $\langle
|G|^{2}\rangle_{\rm B}$ strongly compete, as in the case of $\beta$-(BDA-TTP)$_{2}$SbF$_{6}$. 
The strong competition between $\langle U^{2}\chi_{\rm s}\rangle_{\rm B}$ and $\langle
|G|^{2}\rangle_{\rm B}$ is the origin of the nonmonotonic behavior of
the value of $\lambda$.

%%%%%%%%%%
\begin{figure}[hbtp]
 \begin{center}
 \includegraphics[width=5cm]{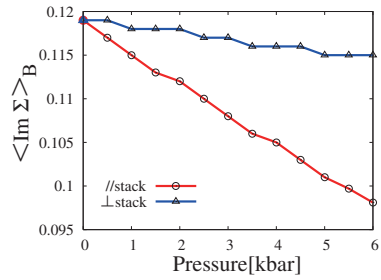}
 \caption{$\langle {\rm
  Im}\Sigma\rangle_{\rm B}$ under uniaxial compression on the two-band model for 
$\beta$-(BDA-TTP)$_{2}$AsF$_{6}$.
}
\label{16}
 \end{center}
\end{figure}%
%%%%%%%%%%
The uniaxial compression dependence of $\langle |G|^{2}\rangle_{\rm B}$ is 
explained by the largest eigenvalue of $\langle|{\rm Im}\Sigma|\rangle_{\rm B}$ as shown in Fig. \ref{16}, 
as in the case of $\beta$-(BDA-TTP)$_{2}$SbF$_{6}$. 
Under uniaxial compression along the both //stack and $\perp$stack directions, the value of 
$\langle|{\rm Im}\Sigma|\rangle_{\rm B}$ decreases monotonically with
increasing pressure.
In particular, along the //stack direction, the value of $\langle|{\rm
Im}\Sigma|\rangle_{\rm B}$ decreases rapidly. This decrease in the value of $\langle|{\rm
Im}\Sigma|\rangle_{\rm B}$ increases the value of 
$\langle |G|^{2}\rangle_{\rm B}$.

\section{Discussion}

As a summary of the results, we show Table II that indicates the suitable
model for explaining the experimental results of $T_c$, mainly focusing on the reproducibility of the 
experimental $T_c$ in a low pressure range, where the linearity of the 
uniform strain seems to be a good approximation. 
The strength of dimerization is estimated by the dimerization factor
($t_{\rm p1}$/$t_{\rm p2}$) because the direction of dimerization corresponds to 
the direction of $t_{p1}$ (intradimer) and $t_{p2}$ (interdimer).

%%%%%%%%%%
\begin{table*}[tb]
\begin{center}
\caption{Relationship between the dimerization factor ($t_{\rm p1}$/$t_{\rm
 p2}$) and the suitable model for the three $\beta$-type salts.}
%\scalebox{0.65}[0.65]{
\begin{tabular}{|c|c|c|c|} \hline
 & $\beta$-(BEDT-TTF)$_{2}$I$_{3}$ & $\beta$-(BDA-TTP)$_{2}$SbF$_{6}$ & $\beta$-(BDA-TTP)$_{2}$AsF$_{6}$ \\ \hline
dimerization factor ($t_{\rm p1}$/$t_{\rm p2}$)
& 2.98 & 2.30 & 2.00  \\ \hline
dimer model 
& $\bigcirc$ & $\bigcirc$ & {\Large$\times$} \\ \hline
2 band model 
& {\Large$\times$} & $\bigcirc$ & $\bigcirc$ \\ \hline
\end{tabular}
\end{center}
\label{cal_exp}
\end{table*}
%\twocolumn
%%%%%%%%%%

Table II shows the following results.
(i) In the case of $\beta$-(BEDT-TTF)$_{2}$I$_{3}$, the low-pressure
behavior of $T_c$ cannot be reproduced by the two-band model, but it can
be
reproduced by the dimer model.
(ii) In the case of $\beta$-(BDA-TTP)$_{2}$SbF$_{6}$, the low-pressure behavior of
$T_c$ can be reproduced by both the dimer model and the two-band
model. We also observe that the dimer model is more suitable than the two-band
model, because the maximum of $T_c$ can be reproduced by the dimer model.
(iii) In the case of $\beta$-(BDA-TTP)$_{2}$AsF$_{6}$, the low-pressure behavior of
$T_c$ cannot be reproduced by the dimer model, but it can be reproduced by the two-band
model. We also observe that the two-band model can reproduce the maximum of $T_c$.
Thus, when the dimerization factor ($t_{\rm p1}$/$t_{\rm p2}$) is large, which 
corresponds to the case of strong dimerization, experimental results are well-reproduced 
by the dimer model. On the other hand, when the dimerization factor is small, the
results of the two-band model are suitable for reproducing the experimental
results. 

One may consider that the results of the dimer model are reproduced by
the two-band model. However, we find that the results of the dimer model
are not always the same
as those of the two-band model with finite $U=10t_{q1}$ because the dimer model corresponds to 
the two-band model with $U\rightarrow\infty$.
The dimer model has the effective $U_{\rm dimer}=2t_{p1}$ at the limit of 
$U\rightarrow\infty$. In fact, we cannot find the value of $U$, using
which the results of $\beta$-(BEDT-TTF)$_{2}$I$_{3}$ in the dimer model
are reproduced in the two-band model.

Our calculation cannot reproduce the experimental results at 
high pressures. Our assumption of the uniform displacements of molecules
at high pressures may be
incorrect. Taking account of the nonlinear distortion
of the lattice against the uniaxial compression is one of the future
problems. Thus, in this paper, our calculation should be compared with the
experimental results obtained at low pressures.

As for the difference among the three materials, we observe that the calculated Fermi
surfaces of $\beta$-(BDA-TTP)$_{2}$SbF$_{6}$ and $\beta$-(BDA-TTP)$_{2}$AsF$_{6}$ 
are slightly thinner than that of $\beta$-(BEDT-TTF)$_{2}$I$_{3}$, which is 
consistent with the recent experimental results of magnetoresistance.\cite{Choi,Yasuzuka}

The structure of the spin susceptibility, which is very sensitive to the
lattice structure, is important for the spin-fluctuation-mediated superconductivity. 
Thus, it may be essential that we employ the hopping integrals obtained by the 
first-principles calculation in order to discuss the
effect of the uniaxial compression in more detail.

\section{Conclusions}

We have studied superconductivity under uniaxial compression in three $\beta$-type 
salts, $\beta$-(BEDT-TTF)$_{2}$I$_{3}$, $\beta$-(BDA-TTP)$_{2}$SbF$_{6}$, and
$\beta$-(BDA-TTP)$_{2}$AsF$_{6}$, theoretically using both the two-band and dimer Hubbard models with FLEX approximation.

We have found that the behavior of $T_c$ in $\beta$-(BEDT-TTF)$_{2}$I$_{3}$ with 
a stronger dimerization is reproduced by the dimer model, while that in weekly dimerized 
$\beta$-BDA-TTP salts is reproduced by the two-band model.
The present study will serve to clarify whether the dimerization of molecules is 
a prerequisite for high $T_{\rm C}$ in organic superconductors. 

Although we apply the uniaxial compression monotonically, 
the shift of $T_c$ exhibits a nonmonotonic behavior against the
pressure. This result is theoretically considered to be due to the competition between the spin frustration 
and the effect of self-energy induced by the spin fluctuation.

\begin{acknowledgments}
This work was supported by a Grant-in-Aid from MEXT, Japan,
for Scientific Research (Grant
No.22540365) from the Ministry of Education,
Culture, Sports, Science and Technology of Japan.
Numerical calculations were performed at the Computer Center 
and the ISSP Supercomputer Center of the University
 of Tokyo, and at the Research Center for Computational Science, Okazaki, Japan.
\end{acknowledgments}

%\bibliographystyle{plain}
%\bibliography{paper1}

\end{document}